\documentclass[conference]{IEEEtran}

\usepackage{cite}
\usepackage{styles/nachiket}
\usepackage{styles/tightlist}
\usepackage{styles/tightenum}

\usepackage[underline=false]{styles/pgf-umlsd}      

\definecolor{light-gray}{gray}{0.8}
\usepackage{wrapfig}
\usepackage{pmboxdraw}
\usepackage[percent]{overpic}
\usepackage[hidelinks,bookmarks=false]{hyperref}
\definecolor{ldx}{RGB}{42,183,202}
\definecolor{cmu}{RGB}{255,162,86}
\definecolor{ic}{RGB}{151,164,106}
\definecolor{bpx}{RGB}{36,42,20}
\definecolor{hl}{RGB}{100,104,109}
\newcommand{\circledat}[2]{%
\node[shape=circle,draw,fill=red,text=white,inner sep=2pt] (char) at (#2) {#1};}

\begin{document}
\title{RapidLayout: Fast Hard Block Placement \\of FPGA-optimized Systolic Arrays 
\\ using Evolutionary Algorithms}


\author{
\IEEEauthorblockN{Niansong Zhang}
\IEEEauthorblockA{Sun Yat-sen University\\
Guangzhou, China\\
{zhangns@mail2.sysu.edu.cn}}
\and
\IEEEauthorblockN{Xiang Chen}
\IEEEauthorblockA{Sun Yat-sen University\\
Guangzhou, China\\
{chenxiang@mail.sysu.edu.cn}}
\and
\IEEEauthorblockN{Nachiket Kapre}
\IEEEauthorblockA{University of Waterloo\\
Ontario, Canada\\
{nachiket@uwaterloo.ca}}
}

\maketitle


\begin{abstract}

Evolutionary algorithms can outperform conventional placement algorithms such as
simulated annealing, analytical placement as well as manual placement on metrics
such as runtime, wirelength, pipelining cost, and clock frequency when mapping
FPGA hard block intensive designs such as systolic arrays on Xilinx UltraScale+
FPGAs.  For certain hard-block intensive, systolic array accelerator designs,
the commercial-grade Xilinx Vivado CAD tool is unable to provide a legal routing
solution without tedious manual placement
constraints. Instead, we formulate an automatic FPGA placement algorithm for
these hard blocks as a multi-objective optimization problem that targets
wirelength squared and maximum bounding box size metrics.  We build an
end-to-end placement and routing flow called RapidLayout using the Xilinx
RapidWright framework.  RapidLayout runs 5--6$\times$ faster than Vivado with
manual constraints and eliminates the weeks-long effort to generate
placement constraints manually for the hard blocks. We also perform automated
post-placement pipelining of the long wires inside each convolution block to
target 650\,MHz URAM-limited operation. RapidLayout outperforms (1) the
simulated annealer in VPR by 33\% in runtime, 1.9--2.4$\times$ in wirelength, and
3--4$\times$ in bounding box size, while also (2) beating the analytical placer
UTPlaceF by 9.3$\times$ in runtime, 1.8--2.2$\times$ in wirelength, and
2--2.7$\times$ in bounding box size. We employ transfer learning from a base FPGA
device to speed-up placement optimization for similar FPGA devices in the
UltraScale+ family by 11--14$\times$ than learning the placements from
scratch.

\end{abstract}



\section{Introduction}

Modern high-end FPGAs provide high compute density with a heterogeneous mixture
of millions of classic lookup tables and programmable routing network along with
tens of thousands of DSP and RAM hard blocks. These hard blocks offer ASIC-like
density and performance for signal processing functions and on-chip SRAM
access. For example, Xilinx UltraScale+ VU11P is equipped with 960 UltraRAM
blocks, 4032 Block RAM slices, and 9216 DSP48 blocks capable of operating at
650--891 MHz frequencies which are typically unheard of with LUT-only designs.
Furthermore, these hard blocks provide specialized nearest-neighbour
interconnect for high-bandwidth, low-latency \textit{cascade} data movement.
These features make it particularly attractive for building systolic neural
network accelerators such as
CLP~\cite{yongming_systolic-clp_isca2017,yongming_systolic-clp_fpl2016},
Cascades~\cite{nachiket_stc_fpl2019}, and Xilinx
SuperTile~\cite{supertile_fpga19, supertile_fpl17}.

Exploiting the full capacity of FPGA resources including hard blocks at high
clock frequency is challenging. The CLP designs presented
in~\cite{yongming_systolic-clp_isca2017,yongming_systolic-clp_fpl2016} only
operate at 100--170\,MHz on Virtex-7 FPGAs but leave DSPs unused. The Xilinx
SuperTile~\cite{supertile_fpga19, supertile_fpl17} designs run at 720\,MHz, but
leave half of the DSPs unused, and also waste URAM bandwidth by limiting access.
The chip-spanning 650\,MHz 1920$\times$9 systolic array design for the VU11P
FPGA~\cite{nachiket_stc_fpl2019} requires 95\% or more of the hard block
resources but fails to route in commercial-grade Xilinx Vivado run with high
effort due to congestion.  Manual placement constraints are necessary to enable
successful bitstream generation, but this requires weeks of painful
trial-and-error effort and visual cues in the Vivado floorplanner for the correct
setup.  This effort is needed largely due to irregularity and asymmetry of the
columnar DSP and RAM fabric and the complex cascade constraints that must be
obeyed for the systolic data movement architecture.  Once the constraints are
configured, Vivado still needs 5--6 hours of compilation time, making design
iteration long and inefficient.  Furthermore, to ensure high-frequency
operation, it becomes necessary to pipeline long wires in the design. Since
timing analysis must be done post-implementation, we end up either suffering the
long CAD iteration cycles or overprovisioning unnecessary pipelining registers
to avoid the long design times. 


Given this state of affairs with the existing tools, we develop RapidLayout: an
alternative, automated, fast placement approach for hard block designs. 
It is important that such a toolflow addresses the shortcomings of the
manual approach by (1) discovering correct placements quickly without the manual
trial-and-error loop through slow Vivado invocations, (2) encoding the complex
placement restrictions of the data movement within the systolic architecture in
the automated algorithm, (3) providing fast wirelength estimation to permit
rapid objective function evaluation of candidate solutions, and (4) exploiting
design symmetry and overcoming irregularity of the columnar FPGA hard block
architecture. Given this wish list, we used the Xilinx RapidWright framework for
our tool.


At its core, the toolflow is organized around the design of a novel evolutionary
algorithm formulation for hard block placement on the FPGA through
multi-objective optimization of wirelength squared and bounding box metrics.
Given the rapid progress in machine learning tools, there is an opportunity to
revisit conventional CAD algorithms~\cite{jeffdean_ml-chip-design_arxiv2019},
including those in this paper, and attack them with this new toolbox.

The key contributions of this work are listed as follows:

\begin{tightlist}
\item We formulate a novel FPGA placement problem for tens of thousands of hard
  blocks as a multi-objective optimization using evolutionary techniques. 

\item We quantify QoR metrics including runtime, wirelength, bounding box size,
  clock frequency, and pipelining cost for the evolutionary placement algorithms
  NSGA-II and CMA-ES. We compare these metrics against conventional Simulated
  Annealing (SA), Genetic Algorithm (GA), Versatile-Place-and-Route
  (VPR)~\cite{vtr2014}, and the state-of-art analytical placer
  UTPlaceF~\cite{li2017utplacef}.
  
\item We build an end-to-end RapidLayout placement-and-routing toolflow using
  the open-source Xilinx RapidWright framework.

\item We develop the transfer learning process for hard block placement to
  accelerate placement optimization through migrating existing placement from
  base devices to similar devices in the UltraScale+ family (VU3P--VU13P).

\end{tightlist}


\section{Background}

We first discuss the hard block intensive systolic array accelerator optimized
for the Xilinx UltraScale+ FPGAs. Next, we discuss the Xilinx RapidWright
framework for programming FPGAs through a non-RTL design flow.
Then, we describe previous research on FPGA placement algorithms.
Finally, we review the classic NSGA-II algorithm and the state-of-art CMA-ES algorithm
and compare them with previous evolutionary placement efforts.

\subsection{FPGA-optimized Systolic Array Accelerator}

\begin{figure}[t]
	\centering
	\includegraphics[width = 0.475\textwidth]{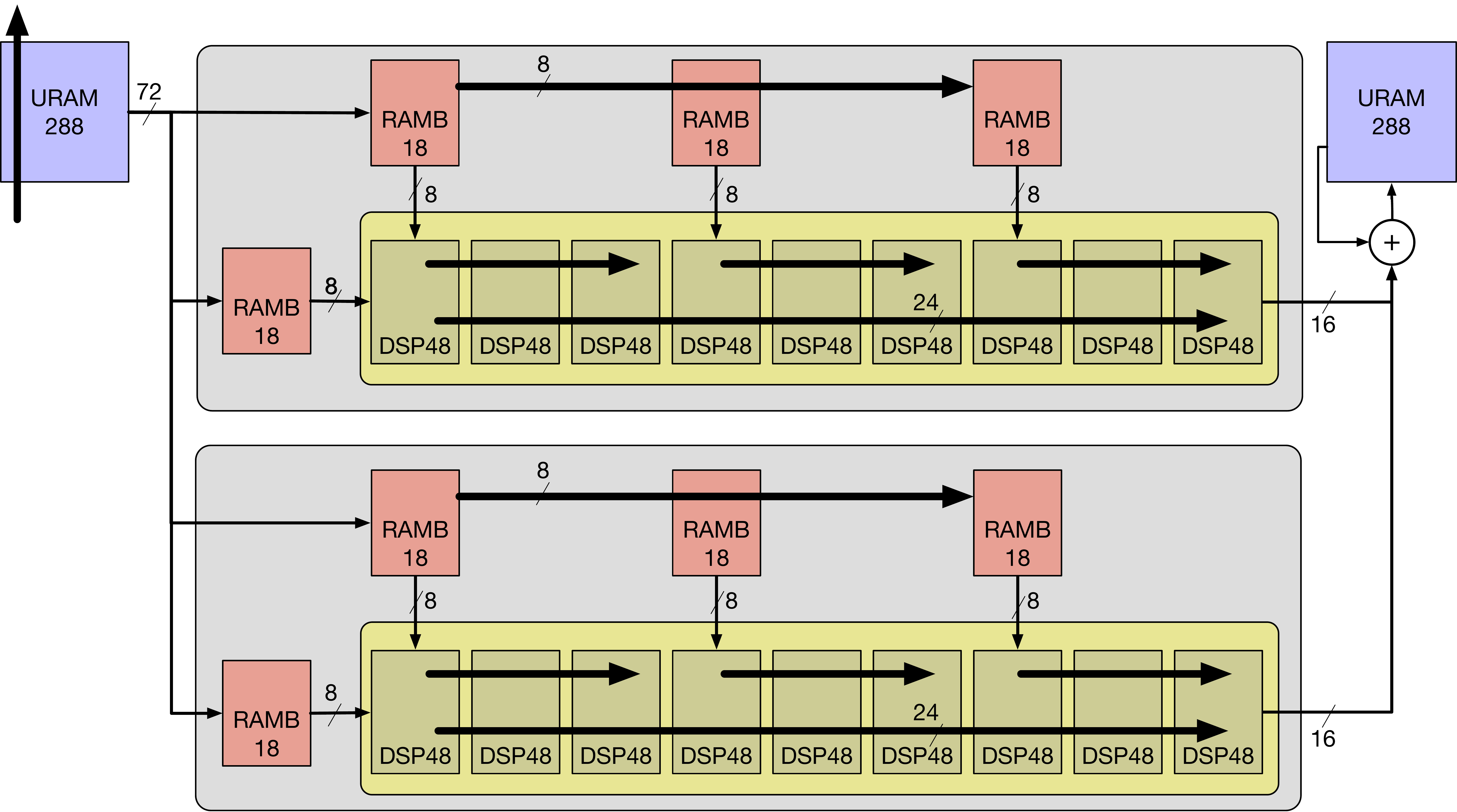}
	\caption{Convolutional Building Block for FPGA-Optimized Systolic Array
  in~\cite{nachiket_stc_fpl2019}. Cascade URAM, BRAM, and DSP links are
highlighted in bold.}
	\label{fig:RTL}
  \vspace{-0.2in}
\end{figure}

Systolic arrays~\cite{jouppi_google-tpu_isca2017, kung_why-systolic_comp1982}
are tailor-made for convolution and matrix operations needed for neural network
acceleration. They are constructed to support extensive data reuse through
nearest-neighbor wiring between a simple 2D array of multiply-accumulate blocks.
They are particularly amenable to implementation on the Xilinx UltraScale+
architecture with cascade nearest-neighbor connections between DSP, BRAM, and
URAM hard blocks. We utilize the systolic convolutional neural network
accelerator presented in \cite{nachiket_stc_fpl2019} and illustrated in
Figure~\ref{fig:RTL}. The key repeating computational block is a convolution engine
optimized for the commonly-used 3$\times$3 convolution operation. This is
implemented across a chain of 9 DSP48 blocks by cascading the accumulators.
Furthermore, row reuse is supported by cascading three BRAMs to supply data to a
set of three DSP48s each. Finally, the URAMs are cascaded to exploit all-to-all
reuse between the input and output channels in one neural network layer.
Overall, when replicated to span the entire FPGA, this architecture uses
95--100\% of the DSP, BRAM, and URAM resources of the high-end UltraScale+ VU37P
device.  When mapped directly using Vivado without any placement constraints,
the router runs out of wiring capacity to fit the connections between these
blocks.  Since the convolution block is replicated multiple times to
generate the complete accelerator, it may appear that placement should be
straightforward. However, due to irregular interleaving of the hard block
columns, and the non-uniform distribution of resources, the placement required to
fit the design is quite tedious and takes weeks of effort.

\subsection{RapidWright}

In this paper, we develop our tool based on the Xilinx
RapidWright~\cite{lavin2018rapidwright} open-source FPGA framework. It aims to
improve FPGA designers' productivity and design QoR (quality of result) by
composing large FPGA designs through a pre-implemented and modular methodology.
RapidWright provides high-level Java API access to low-level Xilinx device
resources. It supports design generation, placement, routing, and allows design
checkpoint (DCP) integration for seamless inter-operability with Xilinx Vivado
CAD tool to support custom flows. 
It also provides access to device geometry information
that enables wirelength calculations crucial for tools that aim to optimize
timing.


\subsection{FPGA Placement}

FPGA placement maps a clustered logical circuit to an array of fixed physical
components to optimize routing area, critical path, power efficiency, and other
metrics.  FPGA placement algorithms can be broadly classified into four
categories: 
(1) classic min-cut partitioning~\cite{trimberger1992placement, maidee2003fast,
maidee2005timing}, 
(2) popular simulated-annealing-based methods~\cite{kirkpatrick1983optimization,
betz1999architecture, betz1997vpr, vtr2014}, 
(3) analytical placement currently used in FPGA CAD tools~\cite{abuowaimer2018gplace3, gort2012analytical,
li2017utplacef, wp416-vivado-user-guide}, and 
(4) esoteric evolutionary approaches~\cite{venkatraman2000evolutionary,
collier2012formal, jamieson2013supergenes}.  
Min-cut algorithm worked well on small FPGA capacities by iteratively
partitioning the circuit to spread the cells across the device.  Simulated
Annealing was the popular choice for placement until recently. It operates by
randomly swapping clusters in an iterative, temperature-controlled fashion
resulting in progressively higher quality results.  Analytical placers are
currently industry standard as they solve the placement problem using a linear
algebraic approach that delivers higher quality solutions with less time than
annealing. For example, Vivado uses an analytical placement to optimize timing,
congestion, and wirelength~\cite{wp416-vivado-user-guide}.

\subsection{Evolutionary Algorithms}
There have been several attempts to deploy evolutionary algorithms for FPGA
placement with limited success. The earliest one by Venkatraman and
Patnaik~\cite{venkatraman2000evolutionary} encodes each two-dimensional block
location in a gene and evaluates the population with a fitness function for
critical path and area-efficiency. More recently, P.
Jamieson~\cite{jamieson2010revisiting}, \cite{jamieson2011exploring} points out that
GAs for FPGA placement are inferior to annealing mainly due to
the crossover operator's weakness and proposed a clustering technique called
supergenes~\cite{jamieson2013supergenes} to improve its performance. 

In this paper, we design a novel combinational gene representation for FPGA hard
block placement and explore two evolutionary algorithms:

\begin{tightenum}
\item {\bf NSGA-II}: Non-Dominated Sorting Genetic Algorithm
  (NSGA-II~\cite{deb2002fast}) is a two-decade-old multi-objective evolutionary
  algorithm that has grown in popularity today for Deep Reinforcement
  Learning~\cite{li2019deep} and Neural Architecture Search~\cite{lu2019nsga}
  applications.  NSGA-II addresses multi-objective selection with non-dominated
  filtering and crowd distance sorting, which allow the algorithm to effectively
  explore the solution space and preserve good candidates.

\item {\bf CMA-ES}: Covariance Matrix Adaptation Evolutionary Strategy (CMA-ES)
  is a continuous domain optimization algorithm for non-linear, ill-conditioned,
  black-box problems \cite{cmapaper}. 
  CMA-ES models
  candidate solutions as samplings of an $n$-dimensional Gaussian variable with
  mean $\mathbf{\mu}$ and covariance matrix $\mathbf{C_{\sigma}}$. At each
  evolutionary iteration, the population is generated by sampling from
  $\mathbb{R}^n$ with updated mean and covariance matrix. 
  Here, crossover and mutation become adding Gaussian noise to the samplings, 
  which overcomes the weakness of GA's crossover operator.
  We use the high-dimensional variant proposed in~\cite{large-scale-cma} for fast
  operation in our placement challenge. 
\end{tightenum}

\begin{figure*}[t]
	\centering
	\resizebox{.5\linewidth}{!}{
	\begin{tikzpicture}[decoration={coil},
dna/.style={decorate, thick, decoration={aspect=0, segment length=0.5cm}}]
 
	\draw[dna, decoration={amplitude=.15cm}] (.1,0) -- (11,0);
	\draw[dna, decoration={amplitude=-.15cm}] (0,0) -- (11,0);
	\node at (5,0.75) {\Large Hard Block Placement Genotype};

	\node [rectangle,rounded corners,inner sep=0pt,minimum width=3cm, minimum
  height=0.25cm,text height=0.95cm, draw=red, ultra thick, fill=red!20,
  anchor=west, opacity=0.4, text
  opacity=1,label=below:{\textcolor{red}{Distribution}}] at (0,0) {};

  \node [rectangle,rounded corners,inner sep=0pt,minimum width=3.8cm, minimum
  height=0.25cm,text height=0.95cm, draw=blue, ultra thick, fill=blue!20,
  anchor=west, opacity=0.4, text
  opacity=1,label=below:{\textcolor{blue}{Location}}] at (3.2,0) {};

	\node [rectangle,rounded corners,inner sep=0pt,minimum width=3.8cm, minimum
  height=0.25cm,text height=0.95cm, draw=olive, ultra thick, fill=olive!20,
  anchor=west, opacity=0.4, text
  opacity=1,label=below:{\textcolor{olive}{Mapping}}] at (7.2,0) {};
 
	\end{tikzpicture}
	
	}
	
	\subfloat[distribution]{
		\includegraphics[width=0.30\textwidth, valign=t]{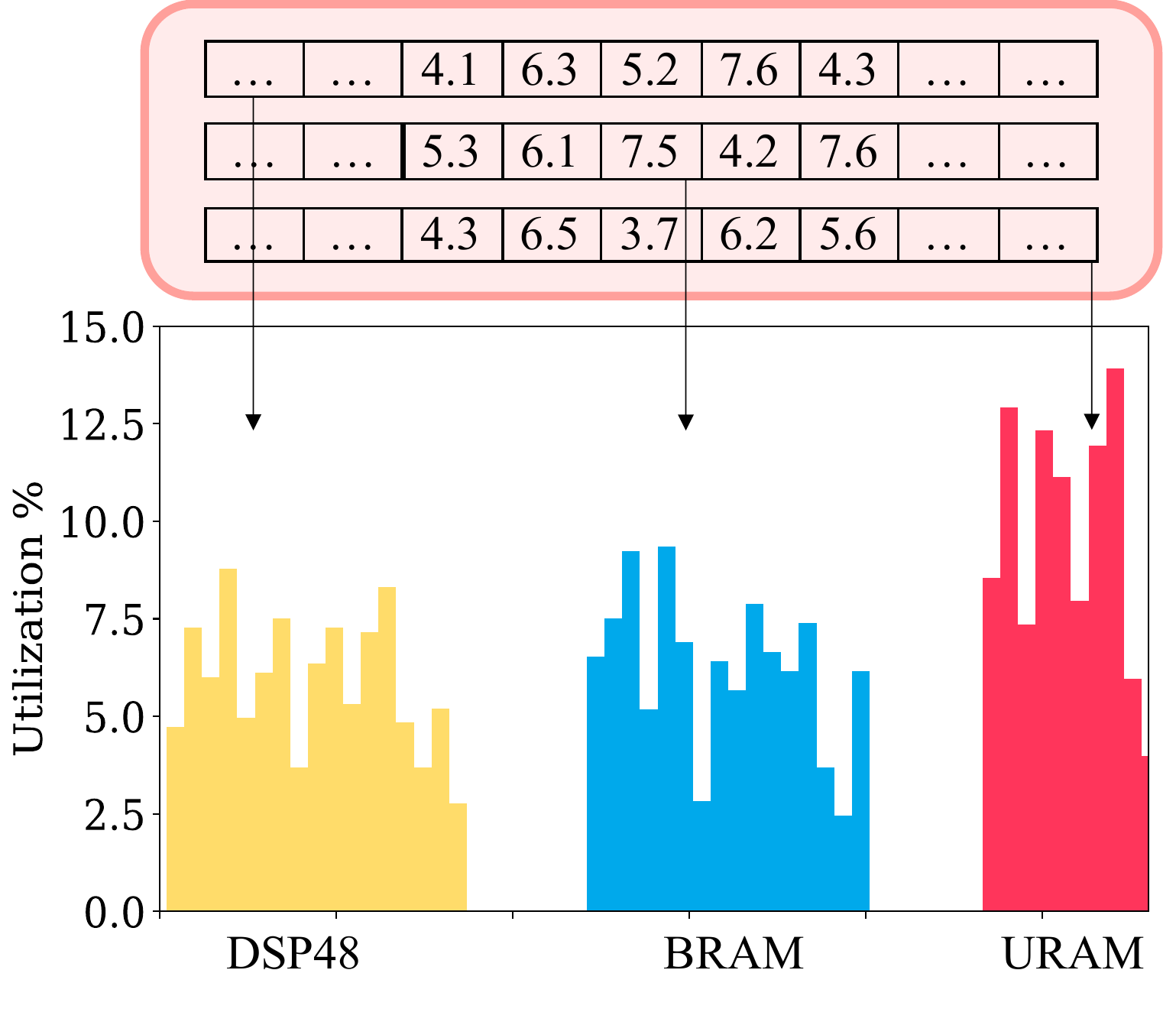}
		\vphantom{
			\includegraphics[width=0.25\textwidth, valign=t]{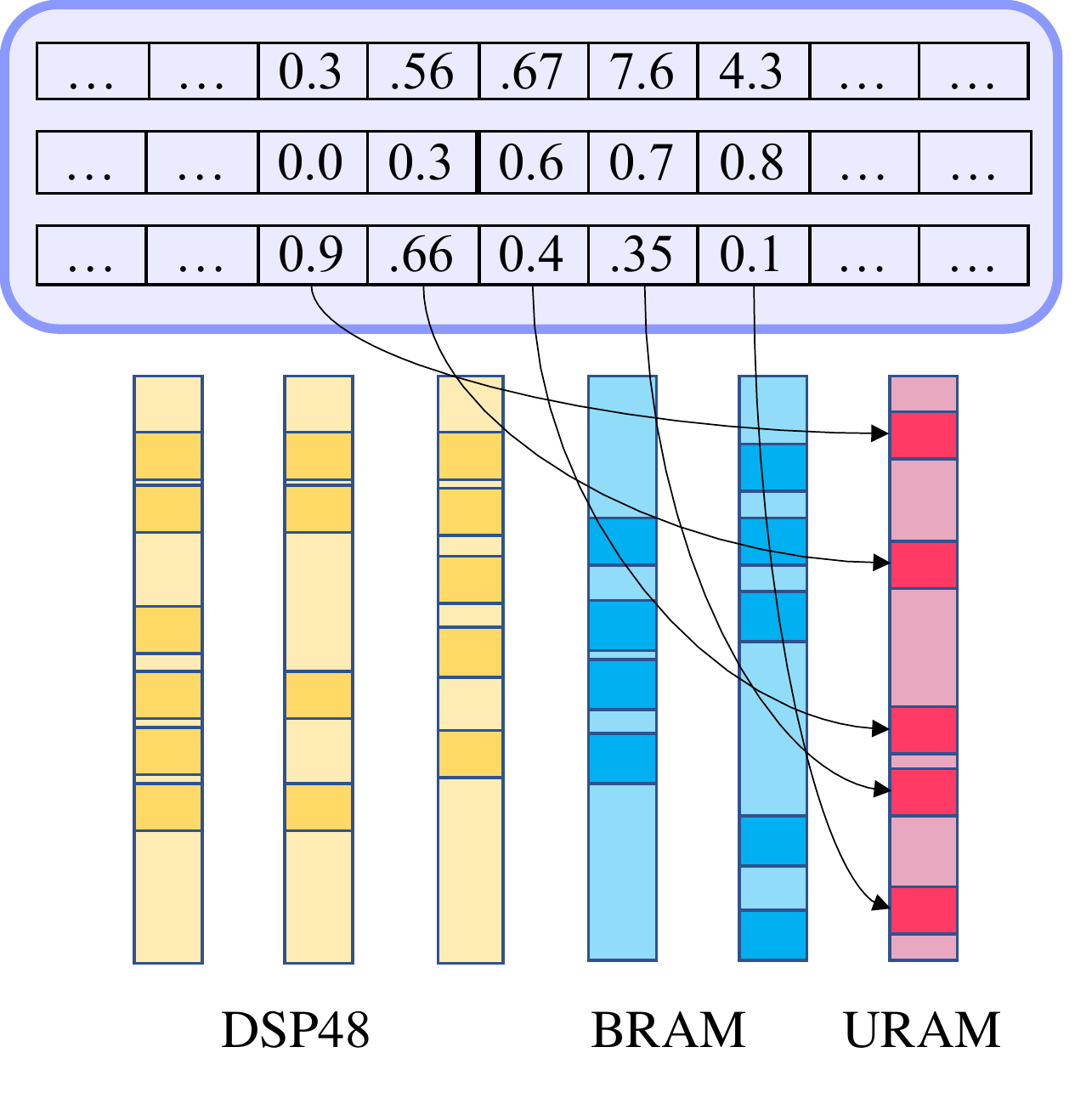}
			\includegraphics[width = 0.23\textwidth, valign=t]{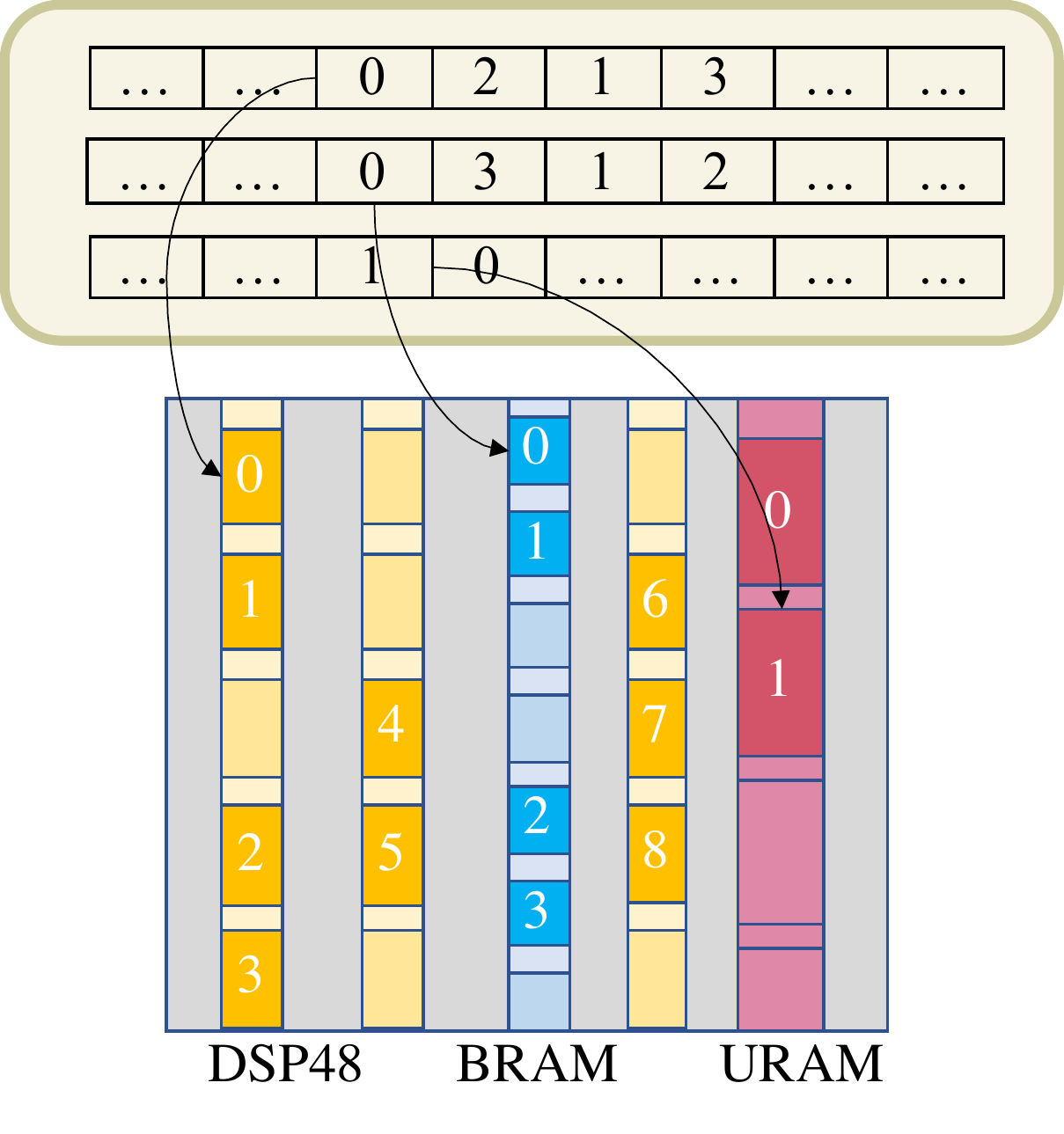}
		}
	}
	\hfill
	\subfloat[location]{
		\includegraphics[width=0.25\textwidth, valign=t]{img/geno-location.pdf}
		\vphantom{
			\includegraphics[width = 0.23\textwidth, valign=t]{img/geno-mapping.pdf}
		}
	}
	\hfill
	\subfloat[mapping]{
		\includegraphics[width = 0.23\textwidth, valign=t]{img/geno-mapping.pdf}
	}
	
	\caption{Our three-tier genotype design for hard block placement. (a) Distribution defines the amount of hard blocks to be placed in each column. (b) Location encodes the relative position of the corresponding hard blocks in its column. (c) Mapping defines the connectivity of hard blocks, i.e., which hard blocks are mapped to one convolution unit. The selected physical hard-block groups are numbered, which corresponds to the mapping genotype.
	}

	\label{fig:genotype}
  \vspace{-0.2in}
\end{figure*}


\section{RapidLayout}

The challenge for mapping FPGA-optimized systolic arrays to the Xilinx UltraScale+ 
device is the placement of hard blocks to their non-uniform, irregular, columnar 
locations on the fabric while obeying the cascade data movement constraints. 
We first present our problem formulation and then discuss how to embed it into
the evolutionary algorithms. 

\subsection{Problem Formulation}

\label{section:formulation}

To tackle the placement challenge, we formulate the coarse-grained placement of 
RAMs and DSP blocks as a constrained multi-objective optimization problem. 
The placement for the rest of the logic \ie lookup tables (LUTs) and flip-flops 
(FFs) is left to Vivado's placer. The multi-objective optimization goal is 
formalized as follows.

\begin{equation} \label{eq:obj1}
	min \sum_{i,j} ((\Delta{x_{i,j}} + \Delta{y_{i,j}}) \cdot w_{i,j})^2 
\end{equation}

\begin{equation} \label{eq:obj2}
	 \min (\max_{k} BBoxSize(C_k))
\end{equation}

subject to:

\begin{equation} \label{eq:region}
  0 \leq x_i,y_i < XMAX,YMAX
\end{equation}

\begin{equation} \label{eq:overlap}
    {x_i,y_i} \neq {x_j,y_j}
\end{equation}

\begin{equation} \label{eq:cascade}
  \begin{aligned}[b]
	& \textrm{if } i \textrm{ is cascaded after } j \textrm{ in the same column: }
  x_i = x_j \\
	& y_i =
	\begin{cases}
		 y_j + 1  & i, j \in \{  DSP, URAM  \} \\
		 y_j + 2  & i, j \in \{  RAMB \}
	\end{cases}
\end{aligned}
\end{equation}

In the equations above: 
\begin{tightlist}
\item $i \in \{ DSP, RAM, URAM \} $ denotes a physical hard block to which a logic
block is mapped. 
\item $C_k$ denotes a convolution unit $k$ that contains 2 URAMs, 18 DSPs, and 8
  BRAMs.
\item $\Delta{x_{i,j}} + \Delta{y_{i,j}}$ is the Manhattan distance between two physical hard blocks $i$ and $j$.
\item $w_{i,j}$ is the weight for wirelength estimation. Here we use the number of connections between hard blocks $i$ and $j$.
\item $BBoxSize()$ is the bounding box rectangle size (width + height) containing the hard
  blocks of a convolution unit $C_k $. 
\item $x_i$ and $y_i$ denote the RPM absolute grid co-ordinates of hard block $i$ that are needed to compute wirelength and bounding box sizes~\cite{ug903}.
\end{tightlist}

{\bf Understanding the Objective Function}: We approximate routing congestion
performance with squared wirelength (Equation~\ref{eq:obj1}) and critical path
length with the maximum bounding box size (Equation~\ref{eq:obj2}). These twin
objectives try to reduce pipelining requirements while maximizing clock
frequency of operation. While these optimization targets may seem odd,
we have observed cases where chasing wirelength$^2$ alone has misled the
optimizer into generating wide bounding boxes for a few stray convolution
blocks. In contrast, optimizing for maximum bounding box alone was observed to
be extremely unstable and causing convergence problems. Hence, we choose these
two objective functions to restrict the spread of programmable fabric routing
resources and reduce the length of critical path between hard blocks and
associated control logic fanout. 

{\bf Understanding Constraints} The optimizer only needs to obey three
constraints. The {\bf region constraint} in Equation~\ref{eq:region} restricts
the set of legal locations for the hard blocks to a particular repeatable
rectangular region of size XMAX$\times$YMAX on the FPGA. The {\bf exclusivity constraint} in
Equation~\ref{eq:overlap} forces the optimizer to prevent multiple hard blocks
from being assigned to the same physical location. The {\bf cascade constraint}
in Equation~\ref{eq:cascade} is the ``uphill'' connectivity restriction imposed
due to the nature of the Xilinx UltraScale+ DSP, BRAM, and URAM cascades. For
DSPs and URAMs, it is sufficient to place connected blocks next to each other.
For BRAMs, the adjacent block of the same type resides at one block away from
the current location. This is because RAMB180 and RAMB181, which are both RAMB18
blocks, are interleaved in the same column. 

\subsubsection{Genotype Design for Evolutionary Algorithms}


We decompose placement into three sub-problems and encode the candidate solutions
with a novel composite genotype design.


\begin{tightenum}
	\item \textcolor{red}{\bf Distribution} Since the systolic array accelerator does not match the hard block resource capacity perfectly, we allocate hard blocks across resource columns according to the distribution genotype.
	\item \textcolor{blue}{\bf Location} Once we choose the exact number of blocks to place on a given resource column, we assign each block in the column a location according to a value between 0 $\rightarrow$ 1. 
	\item \textcolor{olive}{\bf Mapping} Finally, we label selected blocks and allocate them to each convolution unit according to the mapping genotype. It is a permutation genotype that optimizes the order of elements without changing their values.
\end{tightenum}

In Figure~\ref{fig:genotype}, we visualize the genotype design which consists of
the three parts just discussed earlier.  During evolution, each part of the
genotype is updated and decoded independently, but evaluated together.

\subsubsection{Generality of the Evolutionary Formulation}

The problem formulation and genotype design are motivated by a convolutional systolic array. However, the placement formulation 
is not limited to any particular hard block design. For example, the computing unit $C_k$ (Equation~\ref{eq:obj2}) can be any hard block
design, as long as the number of hard blocks, their connections, and cascade information are provided.

\subsubsection{Comparison with Prior Evolutionary Placements}

Our genotype design differs from prior works in three aspects:
(1) We provide placement support for heterogeneous hard block types.
(2) We encode cascade constriants into the genotype, which eschews extra legalization steps and reduces search space. 
(3) The three-tier gentoype design enables placement transfer learning across devices (Section~\ref{section:transfer}).


\tikzstyle{block} = [thick, rectangle, draw, fill=white, text width=12em, text centered, rounded corners, minimum height=2em]
\tikzstyle{invisiblock} = [rectangle, fill=white, text width=12em, text centered, rounded corners, minimum height=2em]
\tikzstyle{decisionblock} = [rectangle, draw, fill=blue!20, text width=12em, text centered, rounded corners, minimum height=2em]
\tikzstyle{line} = [draw, -latex']

\pgfdeclarelayer{bg}    
\pgfsetlayers{bg,main}  

\subsection{RapidLayout Design Flow}

We now describe the end-to-end RapidLayout design flow:

\begin{figure}[h]
  \begin{center}
  \begin{tikzpicture}[node distance=1.25cm and 3.5cm, scale=0.75, transform shape]
    \node [invisiblock] (input) {Convolution Block DCP};
    \node [block, below of=input] (replicate) {Netlist Replication {\bf [\textless 1s]}};
    \circledat{A}{replicate.north west};
    \node [block, below of=replicate, minimum height=7em, node distance=2.15cm] (evolve) {Evolutionary\\ Hard Block\\ Placement\\ {\bf [30\,s--5\,min]}};
    \circledat{B}{evolve.north west};
    \node [block, below of=evolve, node distance=2.2cm] (siteroute) {Placement and\\ Site Routing {\bf [$\approx$3\,min]}};
    \circledat{C}{siteroute.north west};
    \node [block, below of=siteroute] (pipeline) {Post-Placement\\Pipelining {\bf [$\approx$10\,s]}};
    \circledat{D}{pipeline.north west};
    \node [block, below=0.75cm of pipeline] (vivado) {SLR Placement\\ and Routing {\bf [$\approx$55\,min]}};
    \circledat{E}{vivado.north west};
    \node [block, below=0.75cm of vivado] (slr) {SLR Replication {\bf [$\approx$2\,min]}};
    \circledat{F}{slr.north west};
      \node [block, fill=olive!50, right=1cm of evolve] (two) {Compute objective};
    \node [block, fill=olive!50, above of=two] (one) {Generate candidates};
    \node [block, fill=olive!50, below of=two] (three) {Update};
    \begin{pgfonlayer}{bg}
    
    \path[rounded corners, draw=blue, thick, fill=blue!10,opacity=0.9] 
   ([xshift=-0.75em,yshift=0.75em]vivado.north west) to ([xshift=10em,yshift=0.75em]vivado.north east) 
   to ([xshift=10em,yshift=-0.75em]vivado.south east) to ([xshift=-0.75em,yshift=-0.75em]vivado.south west) -- cycle;

    \path[rounded corners, draw=red, thick, fill=red!10,opacity=0.9] 
   ([xshift=-0.75em,yshift=0.75em]replicate.north west) to 
  ([xshift=20em,yshift=0.75em]replicate.north east) to 
  ([xshift=20em,yshift=-0.75em]slr.south east) to
  ([xshift=-0.75em,yshift=-0.75em]slr.south west) to
  ([xshift=-0.75em,yshift=0.75em]slr.north west) to 
  ([xshift=12em,yshift=0.75em]slr.north east) to
   ([xshift=12em,yshift=-0.75em]pipeline.south east) to 
  ([xshift=-0.75em,yshift=-0.75em]pipeline.south west) -- cycle;
    
    \path[rounded corners, draw=olive, thick, fill=olive!10,opacity=0.9] 
   ([xshift=-0.5em,yshift=0.5em]evolve.north west) to ([xshift=0.5em,yshift=0.5em]evolve.north east) 
   to ([xshift=-0.5em,yshift=1.5em]one.north west) to ([xshift=3.5em,yshift=1.5em]one.north east) 
   to ([xshift=3.5em,yshift=-1.5em]three.south east) to ([xshift=-0.5em,yshift=-1.5em]three.south west) 
   to ([xshift=0.5em,yshift=-0.5em]evolve.south east) to ([xshift=-0.5em,yshift=-0.5em]evolve.south west) -- cycle;
    \end{pgfonlayer}
    
    \node [draw=none,right=3.5cm of slr] (RapidWright) {\Large \bf RapidWright};
    \node [draw=none,right=1.75cm of vivado] (RapidWright) {\Large \bf Vivado};
    
    \node [invisiblock,below of=slr] (output) {Bitstream};
    \path [line, thick] (input) -- (replicate);
    \path [line, thick] (replicate) -- (evolve);
    \path [line, thick] (evolve) -- (siteroute);
    \path [line, thick] (siteroute) -- (pipeline);
    \path [line, thick] (pipeline) -- (vivado);
    \path [line, thick] (vivado) -- (slr);
    \path [line, thick] (slr) -- (output);
    \path [line, thick] (one) -- (two);
    \path [line, thick] (two) -- (three);
    \path [line, thick] (three.east) -- ([xshift=0.8cm]three.east) -- ([xshift=0.8cm]one.east) node [midway, above, sloped, rotate=180] (textnode0) {evolve} -- (one.east);
 
  \end{tikzpicture}
  \end{center}
   \caption{RapidLayout Design Flow with runtime details for the Xilinx VU11P
  FPGA along with tool usage information. Bulk of the intelligent exploration is done in RapidWright, and Vivado is only invoked at the end for final placement and routing.}
   \label{fig:flow}
   \vspace{-0.2in}
\end{figure}
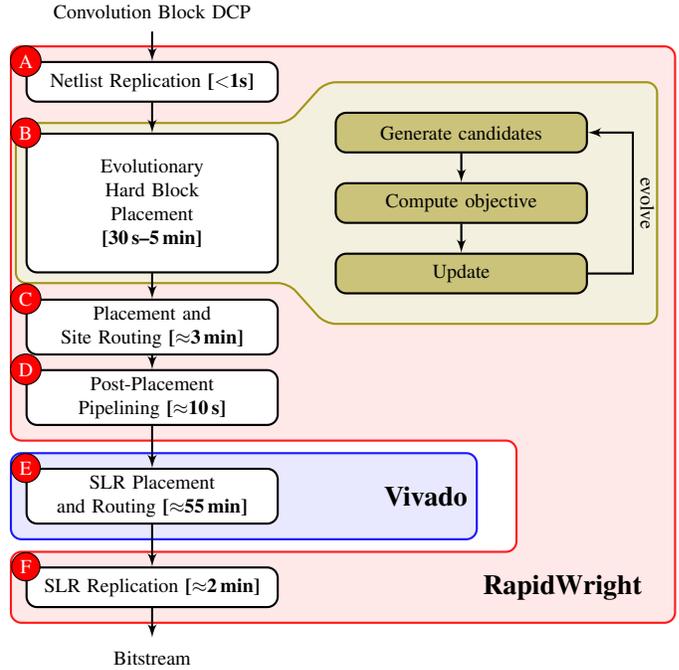


\begin{tightlist}
  \item {\bf Netlist Replication (\textless 1\,s)} RapidLayout starts with a
    synthesized netlist of the convolution unit with direct instantiations of
    the FPGA hard blocks. The unit design is replicated into the entire logical
    netlist that maps to the whole Super Logic Region (SLR).

  \item {\bf Evolutionary Hard Block Placement (30\,s-5min) } RapidLayout uses
    NSGA-II or CMA-ES to generate hard block placement for the minimum repeating
    rectangular region.  Then the rectangular layout is replicated
    (\textit{copy-paste}) to produce the placement solution for the entire SLR. 
    
  \item {\bf Placement and Site Routing ($\approx$3\,min) } The placement
    information is embedded in the DCP netlist by placing the hard blocks on the
    physical blocks called ``sites'', followed by ``site-routing'' to connect
    intra-site pins.
 
  \item {\bf Post-Placement Pipelining ($\approx$10\,s) } After finalizing placement,
    we can compute the wirelength for each net in the design and determine the
    amount of pipelining required for high-frequency operation. This is done
    post-placement~\cite{waver_post-place-retime_fpga2003,eguro_retime-place_fccm2008,desh_integrated-retime-place_fpga2002}
    to ensure the correct nets are pipelined and to the right extent. The
    objective of this step is to aim for the 650\,MHz URAM-limited operation as
    dictated by the architectural constraints of the systolic
    array~\cite{nachiket_stc_fpl2019}. 
  \item {\bf SLR Placement and Routing ($\approx$55\,min) } Once the hard blocks
    are placed and pipelined, we call Vivado to complete LUT/FF placement and
    routing for the SLR.
  \item {\bf SLR Replication (1-5\,min) } The routed design
    on the SLR is copied across the entire device using RapidWright APIs to
    complete the overall implementation. 
\end{tightlist}

For VU11P device, RapidLayout accelerates the end-to-end implementation
by $\approx$5--6$\times$ when measuring CAD runtime alone ($\approx$one
hour vs. Vivado's 5--6 hours). 
This does not include the weeks of manual tuning effort that is avoided by
automatically discovering the best placement for the design.

\subsection{Example Walkthrough}

To illustrate how the different steps help produce the full-chip layout, we walk
you through the intermediate stages of the flow. We will inspect three stages of placement, going from a single block layout to a repeating rectangle layout, and then onto a full-chip layout. 

\begin{figure}[h!]
  \centering
  \includegraphics[width=0.4\textwidth, valign=c]{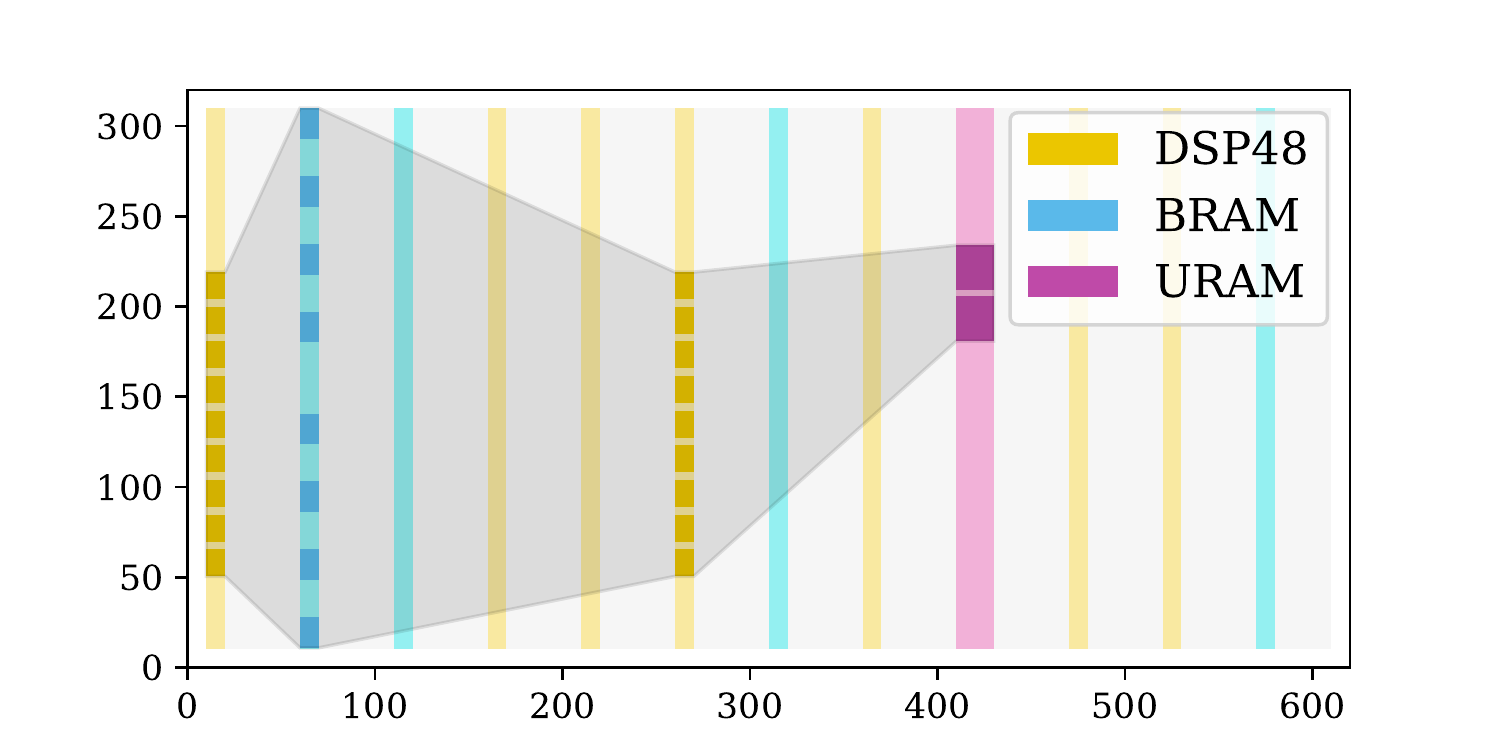}
  \caption{Floorplan layout visualization of a single convolution block
  implementation supporting dual 3$\times$3 kernels to match URAM bandwidth. This
  is the design shown in Figure~\ref{fig:RTL} earlier. The
  bounding polygon that encloses all hard blocks and the routing connections is
  shown in gray.}
  \label{fig:block1}
  \vspace{-0.2in}	
  \end{figure}

{\bf Single Block layout}: The floorplan of a single convolution block in isolation is shown in Figure~\ref{fig:block1}, where the hard block columns and chosen resources are highlighted. 
We also highlight the extent of routing requirements between the hard blocks in gray. 
The locations of the URAM, BRAM, and DSP columns are irregular, which forces a particular arrangement and 
selection of resources to minimize wirelength and bounding box size. 
It is clear that a simple \textit{copy-paste} of a single block is not workable due to this irregularity. 

\begin{figure}[h!]
  \centering
  \includegraphics[width=0.5\textwidth, height=0.2\textwidth, valign=c]{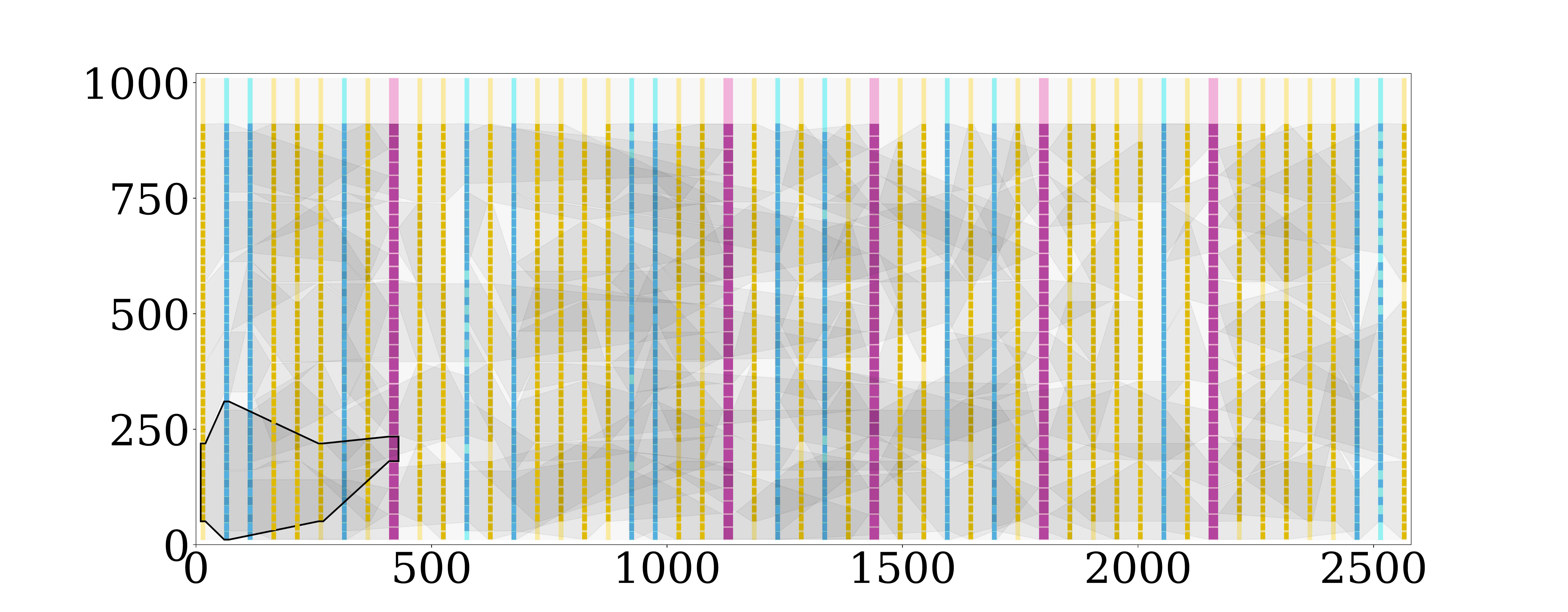}
  \caption{Floorplan layout visualization of a single repeating rectangular region
   layout with 80 convolution blocks. The bounding polygon from
  Figure~\ref{fig:block1} is also shown here for scale.}
  \label{fig:block80}	
  \vspace{-0.1in}
  \end{figure}

{\bf Single Repeating Rectangle Layout}: %
RapidLayout iteratively partitions one SLR and calculates utilizaiton until the divided section does not fit any unit. 
Then, the partition scheme with the highest utilization rate is selected to determine the repeating rectangular region.
In Figure~\ref{fig:block80}, we show the floorplan for such a region. 
The resource utilization within the rectangle is 100\% URAMs, 93.7\% DSP48s, and 95.2\% BRAMs, 
which holds for the entire chip after replication.
Our optimizer minimizes overlapping and thus reducing routing congestions to permit high-frequency operation.

{\bf Full-Chip Layout}: The entire chip layout is generated in two steps: (1) First, 
the rectangular region's placement is replicated to fill up one SLR (SLR0). The SLR is then
pipelined and fully routed. (2) Second, the placement and routing from SLR0's implementation
are replicated across the two other SLRs to fill up the FPGA.

\begin{figure}[h!]
\centering
 \begin{tikzpicture}
   \pgftext{%
     \includegraphics[width=0.23\textwidth]{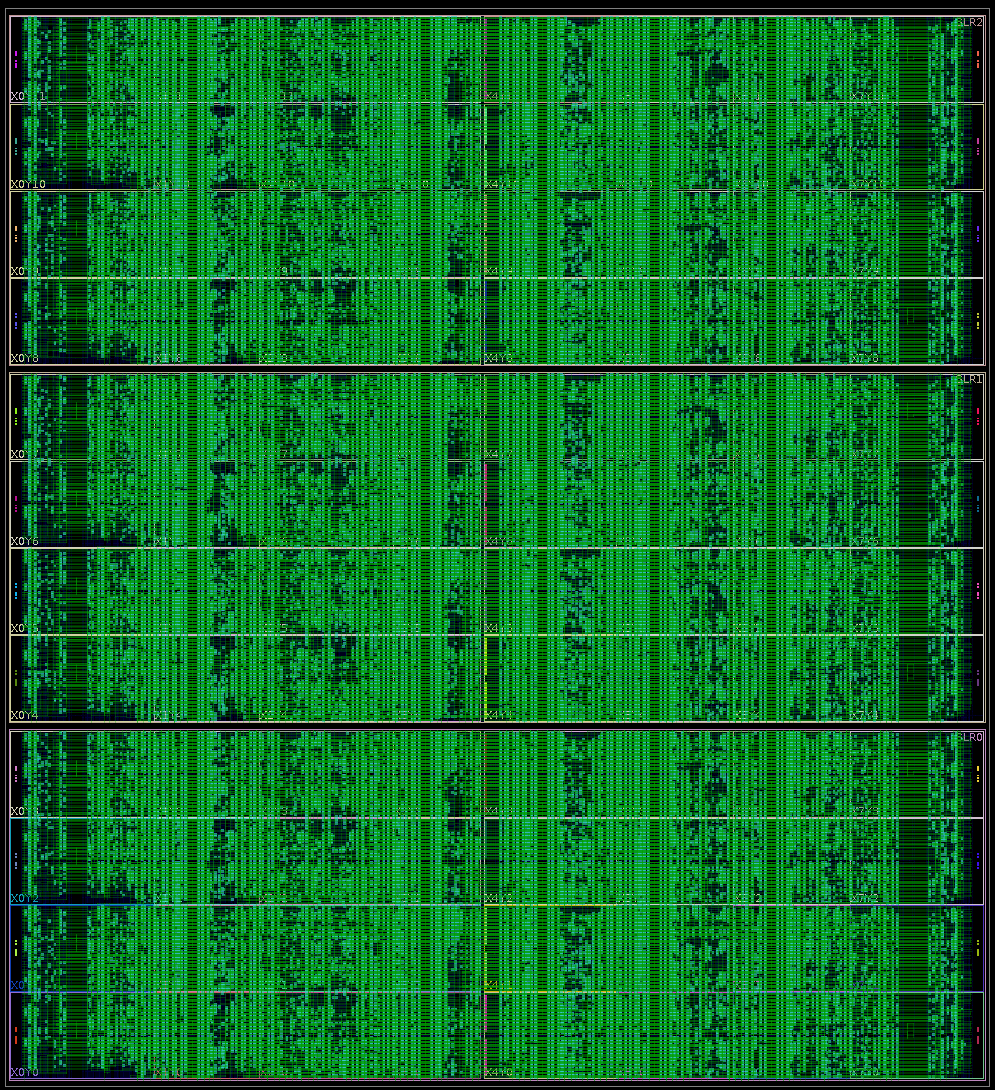}
   }%
   \node [rectangle,rounded corners,minimum width=4cm,minimum
   height=0.7cm,draw=LimeGreen,ultra thick,anchor=south west] (rect1) at
   (0.1-2.2,0.75-2.25) {};
   \node [rectangle,rounded corners,minimum width=4.2cm,minimum
   height=1.5cm,draw=red,ultra thick,anchor=south west,
   text=white,text opacity=0.5] (slr0) at
   (0-2.2,0-2.25) {\Huge SLR0};
   \node [rectangle,rounded corners,minimum width=4cm,minimum
   height=0.7cm,fill=LimeGreen,opacity=0.5,draw
   opacity=1,draw=LimeGreen,ultra thick,anchor=south
   west,text=black,text opacity=1,align=center] (rect0) at
   (0.1-2.2,-2.25+0.05) {Repeating Rect. (Fig~\ref{fig:block80})};
   \node [rectangle,rounded corners,minimum width=4.2cm,minimum
   height=1.5cm,draw=red,ultra thick,anchor=south west,
   text=white,text opacity=0.5] (slr1) at
   (0-2.2,1.5-2.25) {\Huge SLR1};
   \node [rectangle,rounded corners,minimum width=4.2cm,minimum
   height=1.5cm,draw=red,ultra thick,anchor=south west,
   text=white,text opacity=0.5] (slr2) at
   (0-2.2,3-2.25) {\Huge SLR2};
   \draw [ultra thick,->,LimeGreen] (rect0) to [out=0,in=0,looseness=2] node (b)
 [midway,above=0cm,align=center,rotate=-90] {\textcolor{LimeGreen}{Copy Placements +
 }\\\textcolor{LimeGreen}{Vivado P+R}} (rect1);
 \draw [ultra thick,->,red] (slr0) to [out=180,in=180] node (a)
 [midway,above,align=center,rotate=90] {\textcolor{red}{Copy Placement +
 Routing}\\\textcolor{red}{in RapidWright}} (slr2);
   \circledat{F}{[yshift=1.6cm,xshift=0.35cm]a};
   \circledat{C}{[yshift=0.6cm,xshift=0.75cm]b};
   \circledat{D}{[yshift=0.0cm,xshift=0.75cm]b};
   \circledat{E}{[yshift=-0.6cm,xshift=0.75cm]b};
   \draw [ultra thick,->,red] (slr0) to [out=180,in=180] (slr1);
 \end{tikzpicture}
 \caption{Full-chip layout for the systolic array accelerator generated from a
repeating rectangle of size two clock regions high and the full chip wide. After
one replication we span one SLR region. We place and route this with Vivado,
export DCP, reimport into RapidWright to clone across SLRs.}
 \label{fig:fullchip}
 \vspace{-0.2in}
\end{figure}
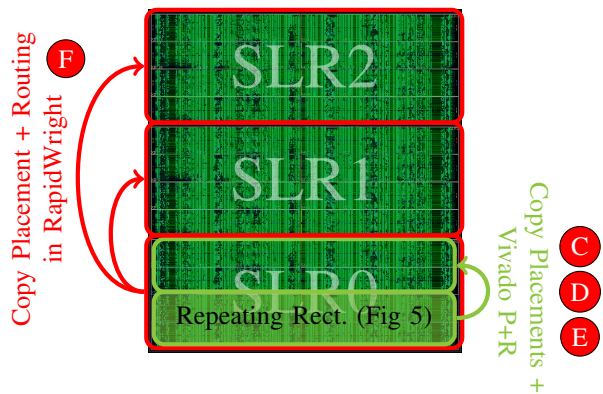

\begin{figure*}[ht]
	\centering
	\subfloat[Wirelength, Bounding Box vs. Runtime Comparison]{
    \label{fig:runtime}
		\includegraphics[width=0.5\textwidth, valign=t]{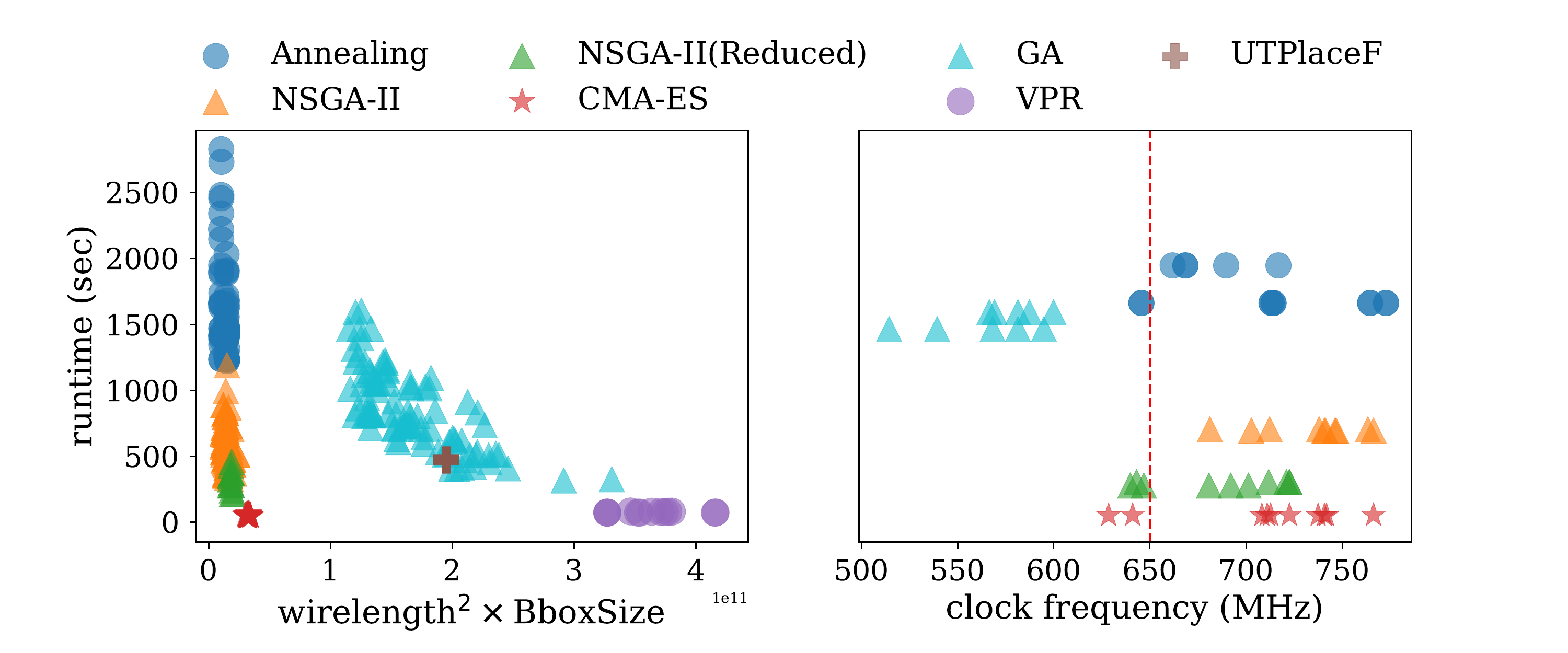}
	}
	\subfloat[Effect of Convergence on Wirelength, Bounding Box]{
    \label{fig:convergence}
		\includegraphics[width=0.5\textwidth, valign=t]{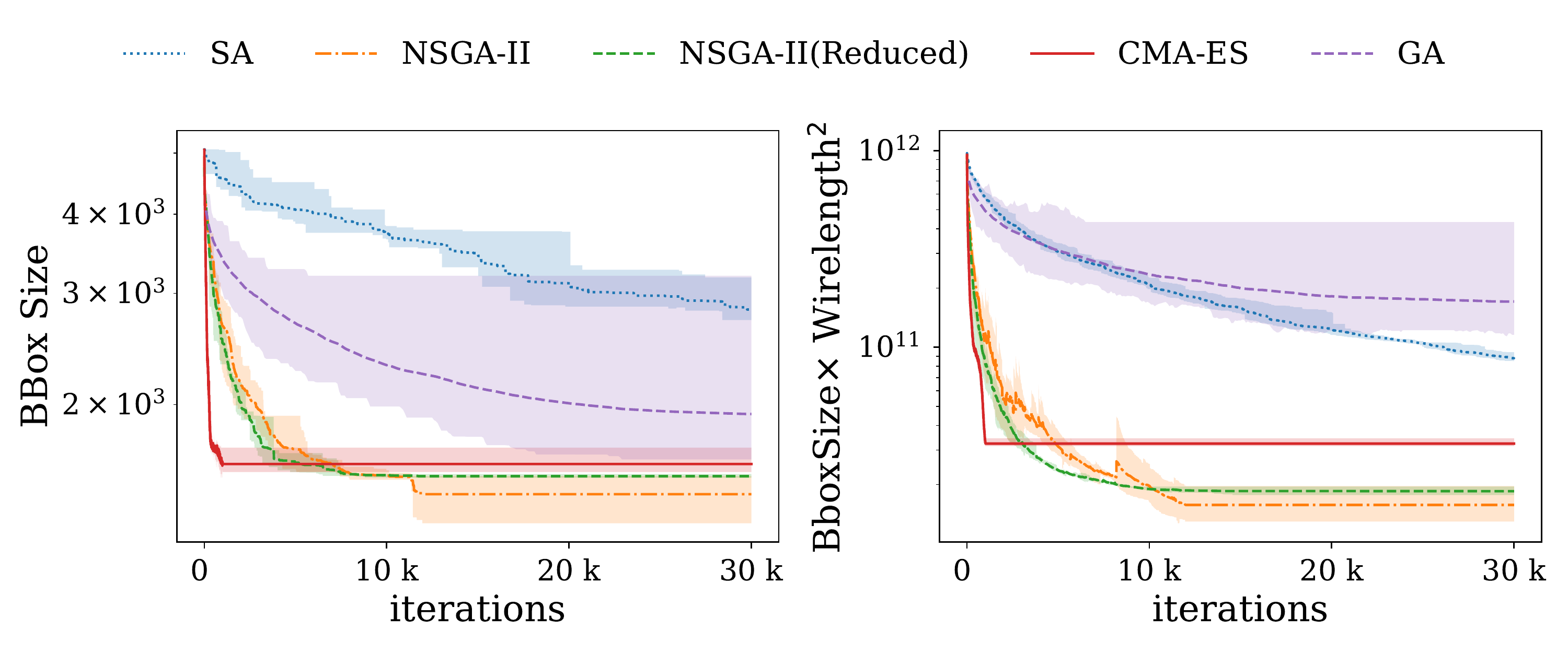}
	}
	
	\caption{Performance, Wirelength, and Bounding Box Comparison: SA, NSGA-II, NSGA-II
    (Red), and CMA-ES
	}
	\label{fig:comparsion}
  \vspace{-0.1in}
\end{figure*}

\begin{table*}[t]
	\caption{Runtime(avg), Wirelength(avg), Max BBox(avg), Pipelining
  Registers(min), and Frequency(avg) for all methods. NSGA-II shows reduced
genotype as well. Speeups and QoR improvements wins by Evolutionary algorithms
also reported in \textcolor{red}{red}$\rightarrow$NSGA-II and \textcolor{olive}{green}$\rightarrow$CMA-ES 
for each competitor algorithm (SA, GA, UTPlaceF, VPR, Manual).}	
	\label{table:comparison}
  \centering
  \begin{adjustbox}{width=\textwidth,center}
  \begin{tabular}{c|c c| c c c c c}
	\toprule
  Method             & NSGA-II    	          & CMA-ES       & SA 					                                                                       & GA 			                                                                  & VPR     	                                                                & UTPlaceF 		                                                                   &  Manual \\
	\midrule
  Runtime (secs)     & 586 (323)		          & 51  		     &1577     (\textcolor{red}{2.7$\times$}, \textcolor{olive}{30.8$\times$})		        & 850 (\textcolor{red}{1.5$\times$}, \textcolor{olive}{16.7$\times$})		      & 76 (\textcolor{red}{0.13$\times$}, \textcolor{olive}{1.5$\times$})		    &  473 (\textcolor{red}{0.8$\times$}, \textcolor{olive}{9.3$\times$})            & 1--2 wks\\
  Wirelength  	     & 3.5K  (3.5K)   		    & 4.4K     		 &3.1K 	   (\textcolor{red}{0.9$\times$}, \textcolor{olive}{0.7$\times$})		          & 9.2K  (\textcolor{red}{2.6$\times$}, \textcolor{olive}{2.1$\times$})    		& 8.5K (\textcolor{red}{2.4$\times$}, \textcolor{olive}{1.9$\times$})   		&  7.8K  (\textcolor{red}{2.2$\times$}, \textcolor{olive}{1.8$\times$})          & 8.1K (\textcolor{red}{2.3$\times$}, \textcolor{olive}{1.8$\times$})  \\
  BBox      	       & 1183  (1543)   		    & 1606     		 &1387	   (\textcolor{red}{1.2$\times$}, \textcolor{olive}{0.9$\times$})		          & 1908  (\textcolor{red}{1.6$\times$}, \textcolor{olive}{1.2$\times$})   		  & 4941	(\textcolor{red}{4.1$\times$}, \textcolor{olive}{3.1$\times$})  		&   3218	(\textcolor{red}{2.7$\times$}, \textcolor{olive}{2.0$\times$})         & 1785 (\textcolor{red}{1.5$\times$}, \textcolor{olive}{1.1$\times$})  \\
  Pipeline Reg.		   & 256K  (273K) 		      & 273K	   		 &273K	   (\textcolor{red}{1.1$\times$}, \textcolor{olive}{1$\times$}) 	 	          & 323K	(\textcolor{red}{1.3$\times$}, \textcolor{olive}{1.2$\times$})   		  & -		                                                                      &   -                                                                            & 306K (\textcolor{red}{1.2$\times$}, \textcolor{olive}{1.1$\times$})   \\
  Frequency	(MHz)	 	 & 733  (688)   & 708 	 &711  (\textcolor{red}{1.03$\times$}, \textcolor{olive}{0.99$\times$})		        & 585 	(\textcolor{red}{1.3$\times$}, \textcolor{olive}{1.2$\times$})	& - 	      	                                                              & - 	                                                                           & 693 (\textcolor{red}{1.1$\times$}, \textcolor{olive}{1.02$\times$}) \\
	\bottomrule
	\end{tabular}
  \vspace{-0.2in}
\end{adjustbox}
\end{table*}



\section{Results}\label{results.sec}


RapidLayout is implemented in Java to enable seamless integration with
RapidWright Java APIs. We use the Java library Opt4J \cite{opt4jpaper} as the
optimization framework for NSGA-II, SA, and GA. CMA-ES is
implemented with Apache Commons Math Optimization Library~\cite{math2013commons}
3.4 API. We use VPR 7.0 official release \cite{vtr2014} and UTPlaceF TCAD version \cite{li2017utplacef}
binary for QoR comparison. All placed designs are routed and timed with Vivado Design Suite 2018.3.
We run our experiments on an Ubuntu 16.04 LTS machine with Intel Xeon Gold 5115 CPU (10 cores, 20 threads) and 128\,GB DDR4 RAM.

\subsection{Performance and QoR Comparison}


We compare the performance and QoR of evolutionary algorithms against
(1) conventional simulated annealing (SA),
(2) academic placement tool VPR 7.0, 
(3) state-of-art analytical placer UTPlaceF, 
(4) single-objective genetic algorithm (GA)~\cite{yang2005evolutionary}, and
(5) manual placement design.
We exclude RapidWright's default annealing-based block placer since it does not give feasible
placement solutions. We run each placement algorithm 50 times with seeded random
initialization. Then, we select the top-10 results for each method to route and
report clock frequency.  While we include VPR and UTPlaceF in comparison,
they do not support {\bf cascade} constraints (Equation \ref{eq:cascade}). This
limits our comparison to an approximate check on solution quality and runtime, 
and we are unable to translate the resulting placement to the FPGA directly.


In Figure~\ref{fig:runtime}, we plot total runtime and final optimized
wirelength$^2$ $\times$ maximum bounding box size for the different placement algorithms
along with Vivado-reported frequency results.
We see some clear trends:
(1) NSGA-II is $\approx$2.7$\times$ faster than SA and delivers 1.2$\times$ bounding box 
improvement, but has $\approx$12.9\% longer wirelength. The average clock
frequency of top-10 results is evidently higher than SA as NSGA-II's performance is more stable.
(2) CMA-ES is $\approx$30$\times$ faster than SA. Although the average bounding box size 
($\approx$16\% larger) and wirelength ($\approx$42\% larger) are worse than SA's results, 
CMA-ES achieves a slightly higher average clock frequency at 711 MHz.
(4) An alternate NSGA-II method discussed later in Section~\ref{sec:nsga-red} with a reduced 
search space delivers roughly 5 times shorter runtime than SA, with only 2.8\% clock frequency degradation, 
which is still above the URAM-limited 650\,MHz maximum operating frequency. 

In Figure~\ref{fig:convergence}, we see the convergence rate of the different
algorithms when observing bounding box sizes and the combined objective. NSGA-II clearly
delivers better QoR after 10\,k iterations, while CMA-ES delivers
smaller bounding box sizes within a thousand iterations. Across multiple runs,
bounding box optimization shows a much more noisy behavior with the exception of CMA-ES. 
This makes it (1) tricky to rely solely on bounding box minimization, and (2) suggests a preference
for CMA-ES for containing critical paths within bounding boxes.

Finally, in Table~\ref{table:comparison}, we compare average metric values
across the 50 runs of all methods.
NSGA-II and CMA-ES deliver superior QoR and runtime against UTPlaceF and VPR, and
speeds up runtime by 3--30$\times$ against annealing with a minor loss in QoR.
Table~\ref{table:comparison} also reports the number of registers needed for
the 650\,MHz operations. NSGA-II delivers this with 17k ($\approx$6\%) less registers
against annealing and 50k ($\approx$16\%) less registers against manual placement.
NSGA-II results in Table~\ref{table:comparison} are run in 20 threads. Although CMA-ES runs
in serial, the runtime is $\approx$10$\times$ faster than NSGA-II with a QoR gap.

\subsection{Parameter Tuning for Annealing and NSGA-II}
In this section, we discuss cooling schedule selection for annealing and
optimizations to NSGA-II to explore quality, runtime tradeoffs.

	\subsubsection{Parameter Tuning for SA}	
	
\begin{figure}[h!]
		\centering
		\includegraphics[width=0.25\textwidth]{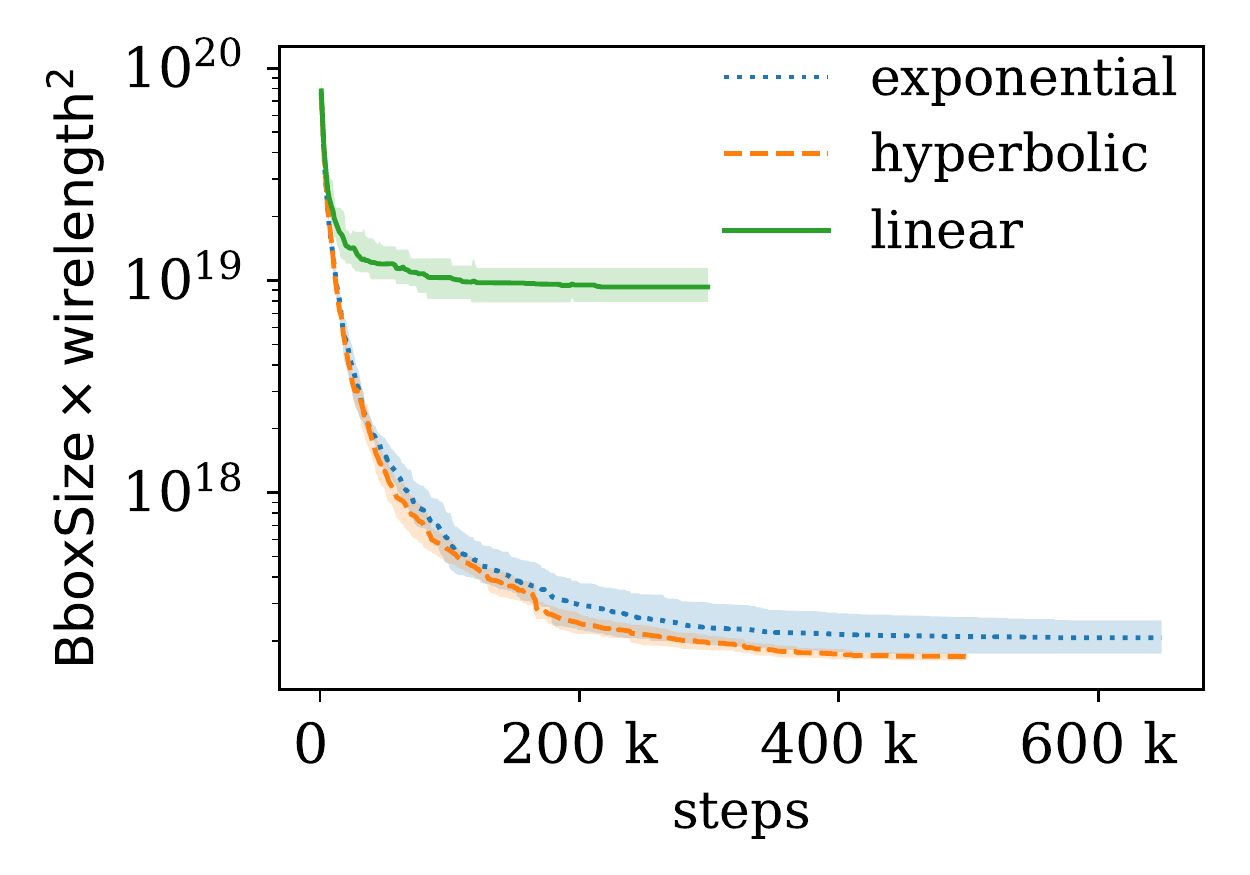}
    \caption{SA Parameter Tuning. Each cooling schedule is experimented with 10 sets of parameters. 
    Annealing placement experiments presented in Table~\ref{table:comparison}
    use hyperbolic cooling schedule for the best QoR performance.}
		\vspace{-0.2in}
		\label{fig:TuningSA}
	\end{figure}

	The cooling schedule determines the final placement quality, but it is highly problem-specific.
	We plot the cooling schedule tuning process in Figure~\ref{fig:TuningSA} and choose the hyperbolic
	cooling schedule for placement experiments presented in Table~\ref{table:comparison} to achieve the best result quality. 

	\subsubsection{NSGA-II Reduced Genotype}
	\label{sec:nsga-red}
	
  As per the genotype design, distribution and location genotypes take up a large
  portion of the composite genotype, and they demand quantization and legalization steps.
  However, for high-utilization designs, distribution and location are less
  influential since resources are nearly fully utilized. Therefore, we reduce
  the genotype to mapping only for NSGA-II, and uniformly distribute and stack
  the hard blocks from bottom to top. As a consequence of this trade-off, we observe
  a $\approx$1.8$\times$ runtime improvement but a 1.3$\times$ larger bounding box size against original NSGA-II. 
  In the convergence plot of Figure~\ref{fig:convergence}, we discover that
  reduced genotype does not save iteration needed, and the bulk of the runtime
  improvements comes from reduced genotype decoding and legalization work.

\subsection{Pipelining}

Finally, we explore the effect of pipelining stages on different placement algorithms. At each 
pipelining depth, multiple placements from each algorithm are routed by Vivado to obtain a frequency
performance range.


\begin{figure}[h!]
	\centering
		\includegraphics[width=0.3\textwidth, valign=c]{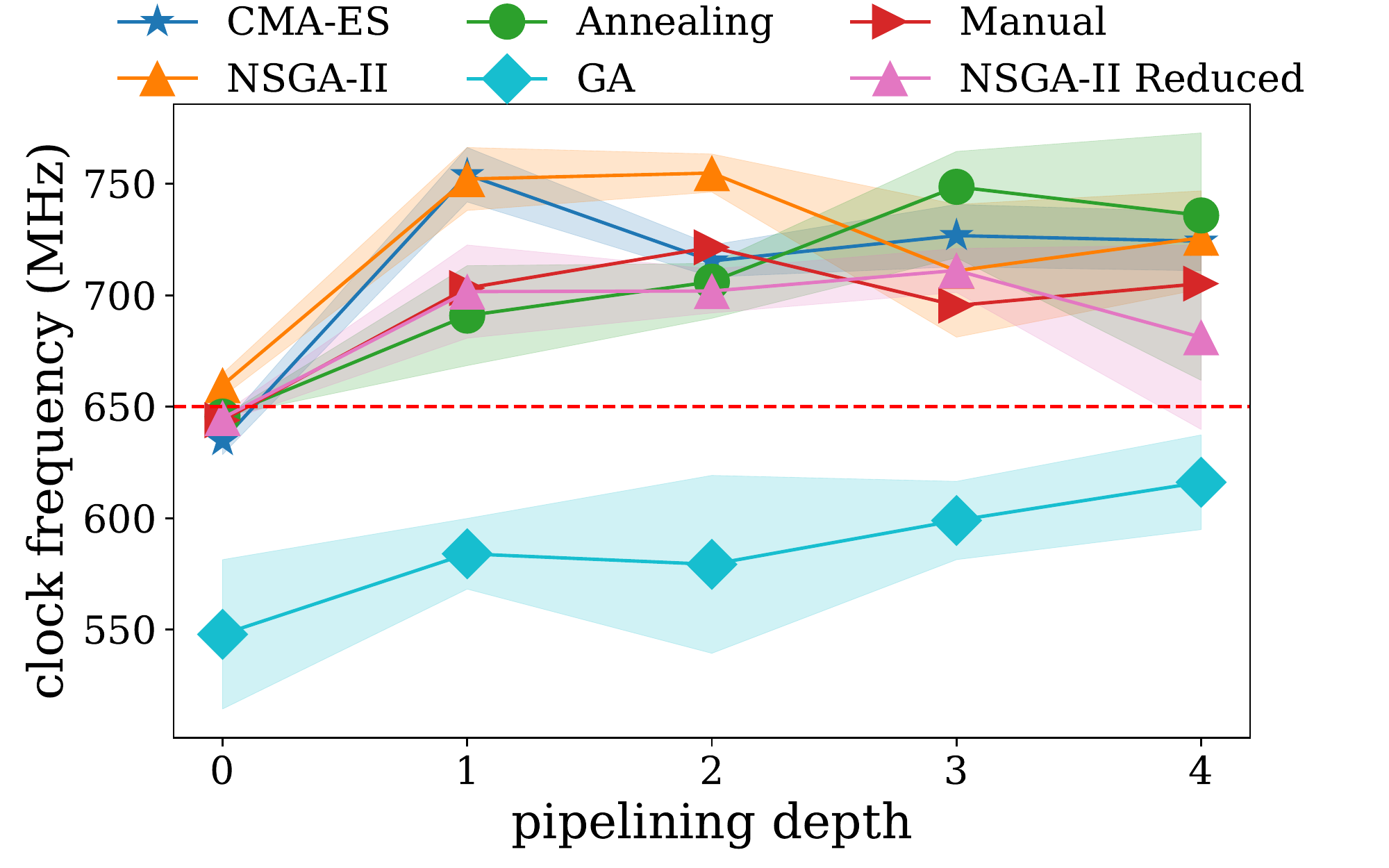}
			
	\caption{Effect of post-placement pipelining on clock frequency of the design.
  NSGA-II delivers 650\,MHz without extra pipelining, while CMA-ES, Annealing,
and Manual placement requires at least one stage. NSGA-II and CMA-ES achieve
750+\,MHz with two stages, while SA requires four stages.}
	\label{fig:pipeline}
  \vspace{-0.1in}
\end{figure}

In Figure~\ref{fig:pipeline}, we show the improvement in frequency as a function
of the number of pipeline stages inserted along the long wires by RapidLayout. We
note that NSGA-II delivers 650\,MHz frequency with no pipelining, while others
require at least one stage. Therefore, NSGA-II saves $\approx$6\%--16\% registers at
pipelining as shown in Table~\ref{table:comparison}. NSGA-II wins over manual
design at every depth, and CMA-ES exhibits the most stable performance. Systolic
array operation at 750+\,MHz should be possible with planned future design
refinements. CMA-ES and NSGA-II can deliver 750+\,MHz frequency with only
two pipeline stages, while SA requires four stages. 




\subsection{Transfer Learning}
\label{section:transfer}

\begin{table}[h!]
  \caption{Transfer Learning Performance: VU3P, VU11P as Seed Devices}
  \label{table:port}
  \centering
  \begin{adjustbox}{width=\columnwidth,center}
  \begin{tabular}{c|c c c c c c}
    \toprule
\multirow{2}{*}{Device}     & Design Size 		& Impl.Runtime 		& \multicolumn{2}{c}{Frequency (MHz)}  	 & \multicolumn{2}{c}{Placement Runtime (s)}    			\\ \cline{4-7}
    			   		          	& (conv units)		& (mins.)	 			 & Scratch & Transfer                & 	Scratch    			& 	Transfer 				      \\
    \midrule
    \emph{xcvu3p}  			    & 123				      &	46.4		       & 718.9	 & 	-		             & 428.3				    & - 							\\
    \emph{xcvu5p}  			    & 246			  	    &   56.9	       & 677.9	 &  660.5	             & 577.9				    & 42.2 	(\textcolor{red}{13.7$\times$})							\\
    \emph{xcvu7p}  			    & 246			        &   55.1	 			 & 670.2	 &  690.1 		             & 578.8				    & 41.9 	(\textcolor{red}{13.8$\times$})							\\
    \emph{xcvu9p}  			    & 369	   			    &   58.4		 	   & 684.9	 &  662.3		             &  570.8				    & 42.0 	(\textcolor{red}{13.6$\times$})							\\ 
    \midrule
    \emph{xcvu11p} 			    & 480	      	    &	65.2           & 655.3   & -	 		             & 522.4				    & -								\\	
    \emph{xcvu13p} 			    & 640        	    & 	69.4       	 & 653.2	 &  701.3                & 443.7				    & 38.4	(\textcolor{red}{11.6$\times$})			\\
    \bottomrule
  \end{tabular}
  \end{adjustbox}
  \vspace{-0.1in}
\end{table}

\begin{figure}[h!]
  \centering
  \includegraphics[width=0.5\textwidth]{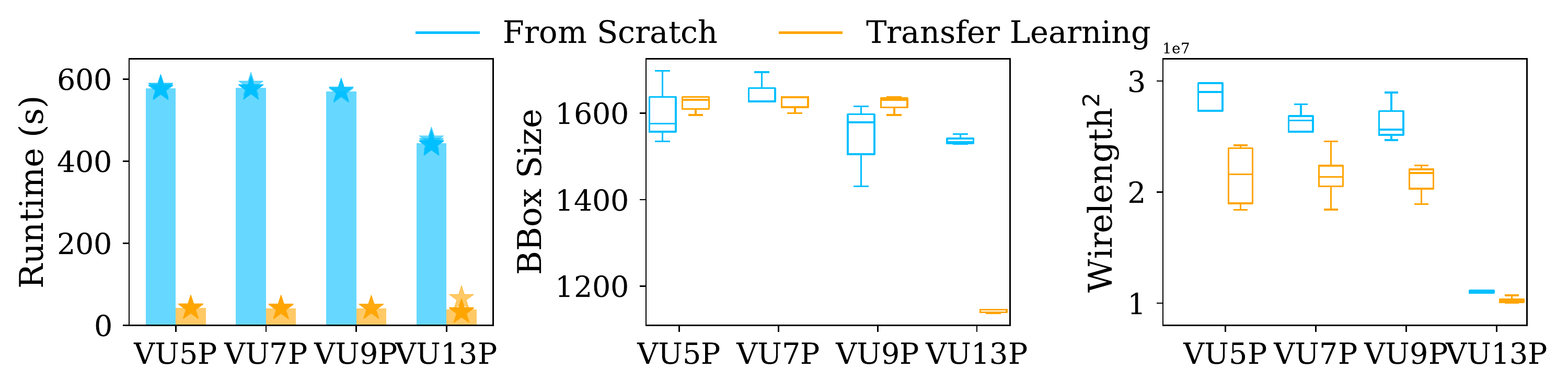}
  \caption{Runtime and QoR comparison between running from scratch and transfer learning. Transfer learning delivers
  $\approx$11--14$\times$ faster runtime, 0.95--1.3$\times$ bbox size improvement, and 1.05--1.17$\times$ wirelength improvement}
  \label{img:transfer}
  \vspace{-0.1in}
\end{figure}

RapidLayout is capable of delivering high-quality placement results on devices
with different sizes, resource ratio, or column arrangements with transfer
learning ability. Transfer learning uses the genotype of an existing placement
as a starting seed for initializing the placement search on a new device.  We partition
Xilinx UltraScale+ family into two groups with a similar number of hard
block columns.  We choose VU3P and VU11P as ``seed'' devices on
which RapidLayout generates placement from scratch with NSGA-II.  Thereafter,
placement results on seed devices are migrated to destination devices in the
same group.  In Table~\ref{table:port} and Figure~\ref{img:transfer}, we compare placement runtimes with and
without transfer learning across a range of FPGA device sizes.  We observe that
transfer learning accelerates the optimization process by 11--14$\times$ with a frequency variation from -2\% to +7\%.
If we observe the total implementation runtime column, we note that SLR
replication ensures that the increase in overall runtime (46
mins.$\rightarrow$69 mins., 1.5$\times$) with device size is much less than the
FPGA capacity increase (123$\rightarrow$640, 5.2$\times$).   




\section{Conclusions}

We present an end-to-end hard block placement workflow for resource-intensive
systolic array designs on modern heterogeneous FPGAs. We show how to outperform
conventional simulated annealing and state-of-art analytical placement with
evolutionary algorithms on metrics such as runtime, bounding box size,
pipelining cost, and clock period. 
RapidLayout also employs transfer learning to quickly generate placements for similar
FPGA devices. RapidLayout is open-source
at \url{https://git.uwaterloo.ca/watcag-public/rapidlayout}.


\section{Acknowledgements}

This work was conducted as part of MITACS Globalink Research Internship in 2019.
This work has been supported in part by NSERC and CMC Microsystems. This work
was also supported by Industry-University Collaborative Education Program between 
SYSU and Digilent Technology: {\it Edge AI Oriented Open Source Software 
and Hardware Makerspace}, and the State's Key Project
of Research and Development Plan in China under Grants 2017YFE0121300-6. 
Any opinions, findings, and conclusions or recommendations expressed in 
this publication are those of the authors and do not necessarily 
reflect the views of the sponsors.

\bibliographystyle{abbrv}
\bibliography{main}

\end{document}